\documentclass[12pt]{article}
\input xy
\xyoption{all}

\usepackage[cp1251]{inputenc}
\usepackage[english]{babel}
\usepackage{amssymb,amsmath}
\usepackage[mathscr]{eucal}
\usepackage{amsfonts}
\usepackage{graphicx}

\begin{document}

\title{\bf
\vspace{1cm} On Gauge Equivalence of Tachyon Solutions in Cubic
Neveu-Schwarz String Field Theory}
\author{
I.Ya. Aref'eva and R.V. Gorbachev\\
\,
 \\
{\it Steklov Mathematical Institute}
\\ {\it Gubkin St.8, 119991, Moscow, Russia.}\\
\\
{\it E-mail:~arefeva@mi.ras.ru,~ rgorbachev@mi.ras.ru} } \maketitle

\abstract{Simple analytic solution to cubic Neveu-Schwarz String
Field Theory including the $GSO(-)$ sector is presented. This solution is an analog of the Erler-Schnabl solution for bosonic case and one of the authors solution for the pure $GSO(+)$ case. Gauge
transformations of the new solution to others known solutions for the $NS$ string
tachyon condensation are constructed explicitly. This gauge equivalence manifestly supports the early observed fact that these solutions have the same value of the action density.}

\baselineskip=16pt

\newpage
\section{Introduction}

The first nontrivial vacuum solution to open string field theory
(SFT) equation of motion has been found in the seminal Schnabl paper
\cite{Sch}. This solution can be presented as a singular limit of a
pure gauge solution configuration \cite{Sch, Okawa}. Schnabl has
found this solution in the Witten bosonic SFT \cite{W}. The similar
solution in the AMZ-PTY super SFT \cite{AMZ, PTY} has been found by
Erler \cite{Erler3}. The solution in fermionic SFT including the
$GSO(-)$ sector \cite{ABKM} has been found in \cite{AGM, AGGKMM}. The Erler
solution $\Phi_E$ and our solution $\widehat\Phi_{AGM}$ both satisfy
to the first and third Sen conjectures. Since both of them are singular limit of the pure
gauge configurations it is tempting to expect that they are related via
a gauge transformation \cite{KF}.

To work with constructed vacuum solutions it would be nice to
perform smoothering gauge transformations. The smooth form of the
vacuum solution to bosonic SFT has been recently found by Schnabl and Erler
\cite{ES}. Following they terminology we call this solution as a
simple analytic one.

The simple analytic solution to cubic SSFT equation of motion is
constructed in \cite{RG}. The goal of this paper is to present a
simple analytic solution to fermionic SFT equation of motion with
non-zero $GSO(-)$ sector. Like to the previous cases \cite{ES, RG}
this new solution involves a continuous integral of wedge state and
no singular limits are necessary. We will show the gauge equivalence
of this solution to the simple pure $GSO(+)$ solution. It was natural
to expect this result, since both solutions have the same value of
action. However this result is not trivial since both solutions are
not pure gauge.

We also present the gauge transformation which relates the new
analytic solution to the AGM analytic solution \cite{AGM}. Note that
the gauge transformation that relats the pure $GSO(+)$ simple solution
$\Phi_G$ and the corresponding solution with phantom terms
\cite{Erler3} has been presented in \cite{RG}.

We have the following picture
$$
\xymatrix{
\Phi_E \ar[r]^{U_{\rm E,G}} & \Phi_G \\
\widehat\Phi_{AGM} \ar[r]^{\widehat U_{\rm AGM,new}} &
\quad\widehat\Phi_{new} }
$$
The first line means that there is a gauge transformation,
$U_{E,G}$, from $\Phi_E$ to $\Phi_G$ and the second line means that
there is a transformation, $\widehat U_{\rm AGM,new}$, from
$\widehat\Phi_{AGM}$ to the solution presented in this paper.

As has been mentioned before, we can construct the gauge
transformation, $\widehat U_{G,new}$, between wo solutions $\Phi_G$ and
$\widehat\Phi_{new}$. This means that we can close the diagram

$$
\xymatrix{
\Phi_E \ar[d]_{\widehat U_{E,AGM}} \ar[r]^{U_{E,G}} & \Phi_G \ar[d]^{\widehat U_{G,new}} \\
\widehat\Phi_{AGM}\quad \ar[r]^{\widehat U_{AGM,new}} &
\quad\widehat\Phi_{new} }
$$
and
\begin{equation}
U_{E,G}\otimes I\cdot\widehat U_{G,new}=\widehat U_{E,AGM}\cdot\widehat
U_{AGM,new}
\end{equation}

Values of the action for the Erler solution $\Phi_E$, the new simple solution $\Phi_G$ and the cubic NS SFT with $GSO(-)$ sector $\widehat\Phi_{AGM}$, were calculated in \cite{Erler3}, \cite{RG, Arroyo:2010fq} and \cite{AGM, AGGKMM}, respectively and gave the same result
\begin{equation}
S[\Phi_E]=S[\Phi_G]=S[\widehat\Phi_{AGM}]=\frac{1}{2\pi^2}.
\end{equation}

\section{Notations}

We use string fields $K,B,c,\gamma$ in split string notation \cite{Erler3, AGM, Erler2},
these fields have the following BRST variations
\begin{equation}
\begin{split}
& Qc=cKc-\gamma^2,\quad Q\gamma=cK\gamma-\frac12\gamma Kc-\frac12\gamma cK,\\
&Q\gamma^2=cK\gamma^2-\gamma^2Kc,\quad
Q B = K,
\end{split}
 \end{equation}
 and satisfy the algebraic relations
\begin{eqnarray}
\{B,c\}=1,&\quad\left[B,\gamma \right] =0,&\quad[c,\gamma]=0,\nonumber\\
{}[B,K]=0,&\quad [K,c]=\partial c,&\quad[K,\gamma]=\partial\gamma,\\
&B^2=c^2=0.&\nonumber
 \end{eqnarray}

Hereinafter we  will use $(1+K)^{-1}$. We rewrite it using the
Schwinger parameterization \cite{ES}
\begin{equation}
\frac{1}{1+K}=\int_{0}^\infty dte^{-t(1+K)}=\int_0^\infty
dte^{-t}\Omega^t.
\end{equation}

\section{New solutions}

Here we present a new solution to the fermionic string field
equations of motion
\begin{equation}
\begin{split}\label{eom}
&Q\Phi_++\Phi_+\star\Phi_+-\Phi_-\star\Phi_-=0,\\
&Q\Phi_-+\Phi_+\star\Phi_--\Phi_-\star\Phi_+=0.
\end{split}
\end{equation}
The simple analytical solution is\footnote{We can also construct the following ill-defined solution \begin{equation*}\begin{split}\Phi_+&=(c+B\gamma^2)(1-K)+B\gamma^2,\\ \Phi_-&=\gamma-\frac12\gamma BcK+\frac12\gamma KBc,\end{split}\end{equation*}
which formally satisfies the equation of motion (\ref{eom}). In the case of the pure $GSO(+)$ sector the ill-defined solution has the form \cite{Arroyo:2010fq}
\begin{equation}
\Phi_+=(c+B\gamma^2)(1-K).
\end{equation}}
\begin{equation}
\begin{split}\label{Phi+-}
\Phi_+&=(c(1+K)Bc+2B\gamma^2)\frac{1}{1+K},\\
\Phi_-&=\left(\gamma+cBK\gamma+\frac12\gamma KBc+\frac12\gamma BcK
\right)\frac{1}{1+K}.
\end{split}
\end{equation}
This solution is a generalization of the solution to the superstring
equation of motion \cite{RG}
\begin{equation}\label{Phi}
\Phi=(c(1+K)Bc+B\gamma^2)\frac{1}{1+K},
\end{equation}
which in part is a generalization of Erler-Schnabl's solution
to bosonic field equation of motion \cite{ES}.

To derive the gauge equivalence (\ref{Phi+-}) and (\ref{Phi}) it is useful to introduce two functions \cite{ES}
\begin{equation}\label{fg}
f=1,\quad g=\frac{1}{1+K},
\end{equation}
and rewrite (\ref{Phi+-}) and (\ref{Phi}) in the form
\begin{equation}\label{newPhi+-}
\begin{split}
\Phi_+&=fc\frac{KB}{1-fg}cg+f\gamma\frac{KB}{1-fg}\gamma g+fB\gamma(1-K)\gamma g,\\
\Phi_-&=fc\frac{KB}{1-fg}\gamma
g+f\gamma\frac{KB}{1-fg}cg-\frac12f\gamma KBcg+\frac12f\gamma BcKg,
\end{split}
\end{equation}
and
\begin{equation}\label{newPhi}
\Phi=fc\frac{KB}{1-fg}cg+fB\gamma^2g.
\end{equation}
If we choose $f=g=F$ we get the AGM solution and the Erler
solution.

\section{Gauge equivalence of solutions}
\subsection{$U_{E,G}$ and $\widehat U_{AGM,new}$}

Following \cite{ES} one can built a gauge transformation
between (\ref{newPhi}) and Erler's solution $\Phi_E$ \cite{RG}
\begin{equation}
\Phi_E=U^{-1}_{E,G}(\Phi_G+Q)U_{E,G},
\end{equation}
where
\begin{equation}
\begin{split}
U_{E,G}&=1-fBcg+Mf'Bcg',\\
U^{-1}_{E,G}&=1-f'Bcg'+M^{-1}fBcg,
\end{split}
\end{equation}
and the function $M$ is defined as
\begin{equation}
M=\left(\frac{1-fg}{1-f'g'}\right),
\end{equation}
here $f'=g'=F$ and $f,g$ are defined by (\ref{fg}).

The gauge transformation between (\ref{newPhi+-}) and the AGM
solution $\widehat\Phi_{AGM}$ is
\begin{equation}\label{gauge}
\widehat\Phi_{AGM}=\widehat U^{-1}_{\rm
AGM,new}(\widehat\Phi_{new}+\widehat Q)\widehat U_{AGM,new}
\end{equation}
where
\begin{equation}
\begin{split}
&\widehat U_{AGM,new}=U_+\otimes I+U_-\otimes\sigma_1,\\
&\widehat U^{-1}_{AGM,new}=U_+^{-1}\otimes
I+U_-^{-1}\otimes\sigma_1,
\end{split}
\end{equation}
and
\begin{equation}
\begin{split}
U_+=(1-fBcg+Mf'Bcg'),&\quad U_-=(-fB\gamma g+Mf'B\gamma g'),\\
U^{-1}_+=(1-f'Bcg'+M^{-1}fBcg),&\quad U^{-1}_-=(-f'B\gamma
g'+M^{-1}fB\gamma g).
\end{split}
\end{equation}
Also we denote
\begin{equation}
\widehat Q=Q\otimes\sigma_3,\quad
\widehat\Phi=\Phi_+\otimes\sigma_3+\Phi_-\otimes i\sigma_2.
\end{equation}

\subsection{$\widehat U_{G,new}$}

Here we build the gauge transformation connecting two solutions:
solution to super and fermionic field theory equation of motion.

Let us consider the gauge transformation
\begin{equation}\label{gauge3}
\widehat\Phi'=\widehat U^{-1}(\widehat\Phi+\widehat Q)\widehat U
\end{equation}
and rewrite it in the components
\begin{equation}
\begin{split}\label{gauge2}
\Phi_+'&=U^{-1}_+(\Phi_++Q)U_+-U^{-1}_-(\Phi_++Q)U_-+U^{-1}_+\Phi_-U_--U_-^{-1}\Phi_-U_+,\\
\Phi_-'&=U^{-1}_+(\Phi_++Q)U_--U^{-1}_-(\Phi_++Q)U_++U^{-1}_+\Phi_-U_+-U_-^{-1}\Phi_-U_-.
\end{split}
\end{equation}
We take $\Phi_+$ and $\Phi_-$ in the form (\ref{newPhi+-}) and take $U_+=U^{-1}_+=I$ and $U_-=-fB\gamma g$, $U_-^{-1}=fB\gamma
g$.

Then the first line of (\ref{gauge2}) gives
\begin{equation}
\begin{split}
&U^{-1}_+(\Phi_++Q)U_+=fc\frac{KB}{1-fg}cg+f\gamma\frac{KB}{1-fg}\gamma
g+fB\gamma(1-K)\gamma g,\\
&U^{-1}_-(\Phi_++Q)U_-=-f\gamma\frac{KB}{1-fg}f^2g^2\gamma g-f\gamma
KBfg\gamma g,\\
&U^{-1}_+\Phi_-U_-=-f\gamma\frac{KB}{1-fg}fg\gamma g,\\
&U_-^{-1}\Phi_-U_+=f\gamma\frac{KB}{1-fg}fg\gamma g.
\end{split}
\end{equation}
So we have the following expression for $\Phi_+'$
\begin{equation}
\Phi_+'=fc\frac{KB}{1-fg}cg+fB\gamma^2g.
\end{equation}
For the second line of (\ref{gauge2}) we have
\begin{equation}
\begin{split}
&U_+^{-1}(\Phi_++Q)U_-=-fc\frac{KB}{1-fg}fg\gamma g-fcKB\gamma g-\frac12f\gamma KBcg-\frac12f\gamma BcKg,\\
&U_-^{-1}\Phi_+U_+=f\gamma\frac{KB}{1-fg}fgcg,\\
&U_+^{-1}\Phi_-U_+=fc\frac{KB}{1-fg}\gamma
g+f\gamma\frac{KB}{1-fg}cg-\frac12f\gamma KBcg+\frac12f\gamma BcKg,\\
&U_-^{-1}\Phi_-U_-=0,
\end{split}
\end{equation}
thereby, for $\Phi_-'$ we have
\begin{equation}
\Phi_-'=0.
\end{equation}
Thus, we have the gauge equivalence of two solutions.

\section{Conclusion}

The construction presented in this paper permits to conclude that
all the tachyon condensation solutions are gauge equivalent. In standard Siegel gauge the tachyon leaves in the $GSO(-)$ sector and breaks the
supersymmetry of the model. It is also can be expected that the analytical
solution \cite{Erler3} leaving in the pure $GSO(+)$ in spirit of year considerations in \cite{AMZ2} breaks the supersymmetry. To
check this statement one has to also deal with the R-sector. It
would be interesting to incorporate the Kroyter idea of constructing
the NSR SFT whose string fields carry an
arbitrary picture number and reside in the large Hilbert space \cite{Kroyter}.

\newpage
\section*{Acknowledgements}
The work is supported in part by RFBR grant 08-01-00798,
NS-795.2008.1 and FCPK-02.740.11.5057.

 \end{document}